\documentclass[12pt]{article}
\usepackage{hyperref}
\usepackage{amsmath}
\usepackage{amsfonts}

\usepackage{graphicx}
\graphicspath{ {./images/} }
\usepackage{changepage,parskip}

\newcommand{\mycomment}[1]{}

\newenvironment{sciabstract}{%
\begin{quote} }
{\end{quote}}

\title{A Methodology to Measure Impacts of  Scenarios Through Expected Credit Losses} 

\author{Mahmood Alaghmandan,$^{1,}$\footnote{Corresponding Author.},
\hspace{1mm}   Meghal Arora$^{2}$,  \hspace{1mm}  Olga Streltchenko$^{3}$\\
\\
\normalsize{\it $^{1}$Manager, Model Validation and Risk Management}\\
\normalsize{Farm Credit Canada,}\\
\normalsize{ 100 Queen St, unit 1460, Ottawa, ON K1P 1J9 Canada}\\
\small{$^{1}${\bf Email:} \texttt{mahmood.alaghmandan@fcc.ca}}\\
\\
\normalsize{\it $^{2}$Principal Analyst, Catastrophic Risk Division}\\
\normalsize{Office of the Superintendent of Financial Institutions,}\\
\normalsize{ 255 Albert Street,Ottawa, Ontario K1A 0H2 Canada}\\
\small{$^{2}${\bf Email:} \texttt{meghal.arora@osfi-bsif.gc.ca},}\\
\\
\normalsize{$^{3}$ Ottawa, ON Canada}\\
\small{{\bf Email:} \texttt{streltch@gmail.com} }
}
\date{\today}

\begin{document} 


\baselineskip24pt


\maketitle 

% Abstract

\begin{sciabstract}
  {\bf Abstract:} {In this paper, we present a methodology for measuring the impact of scenarios on the expected losses of exposures by leveraging the existing provisioning infrastructure within financial institutions, where scenario effects are captured through changes in probabilities of default. We then describe how to design and implement a scenario test where risk drivers are given for standardized groupings of exposures, and the groupings are defined based on common features of the exposures. The methodology presented served as a theoretical foundation for the standardized climate scenario exercise conducted in 2024 by the Office of the Superintendent of Financial Institutions of Canada and Qu\'{e}bec’s  Autorit\'{e} des March\'{e}s Financiers.}
\end{sciabstract}

% Main Body

This paper develops a general framework for evaluating the impact of scenarios\footnote{The framework can be applied to any set of scenarios--macroeconomic or otherwise--that can be translated into expected credit loss risk drivers compatible with provisioning measurements, including but not limited to climate transition or physical risk scenarios. Throughout the paper, any reference to''scenarios'' should be understood in this broad sense.} on the expected losses of a financial portfolio by leveraging existing provisioning measurements, where scenario effects are represented through changes in the probabilities of default of underlying exposures. 

Provisioning is a core risk-management and accounting mechanism through which financial institutions recognize expected credit losses on their balance sheets. Historically, provisions emerged as a prudential response to the inherently uncertain nature of lending, aiming to absorb losses and protect solvency. Their modern evolution, however, was decisively shaped by the 2007–2009 Global Financial Crisis. During the crisis, incurred-loss accounting frameworks proved deeply procyclical: credit losses were recognized too late, only after clear evidence of impairment had materialized, thereby amplifying balance-sheet stress and eroding market confidence precisely when resilience was most needed. In response, regulators and standard setters sought mechanisms that would force earlier recognition of credit deterioration, strengthen capital adequacy, and mitigate the feedback loop between financial distress and the real economy.

This shift gave rise to forward-looking provisioning regimes, most notably lifetime expected credit loss frameworks such as those embedded in both International Accounting Standard Board (IASB) and Financial Accounting Standard Board (FASB) proposed provisioning standards, namely IASB's International Financial Standard Reporting number 9 (IFRS9) and IASB's Current Expected Credit Loss (CECL).  Under these approaches, provisions are no longer backward-looking reflections of realized defaults but dynamic estimates of future losses, capable of measuring it over the expected life of an exposure, conditional on macroeconomic scenarios and borrower credit quality. 

There have been many attempts to leverage provisioning calculations for scenario analysis. For example, Breed et al. \cite{ifrs9} discuss a methodology for embedding macroeconomic forecasts into IFRS 9 PD term structures, using principal component regression to adjust lifetime PD profiles. In an IMF working paper, Gross et al. \cite{imf} present tools for top-down stress testing that are compatible with IFRS 9 and CECL, and that are useful for regulators and supervisors in scenario-conditional ECL estimation. A similar approach to leveraging ECL calculations for climate scenario analysis was employed in the Bank of Canada and Office of the Superintendent of Financial Institutions of Canada (OSFI) Climate Scenario Analysis Pilot \cite{boc}.

In this paper, we propose a methodology for conducting scenario analysis using provisioning calculations. By embedding expectations about future paths of the economy and credit risk over the lifetime of exposures, provisioning provides a natural framework for scenario analysis that incorporates a coherent macro-financial narrative and translates it into credit implications for financial institutions’ obligors.

In addition, we consider the case where risk parameters are stressed at the level of standardized groupings of exposures where the groupings are defined based on common features of the exposures. This approach drastically reduces the number of risk factors and limits capturing idiosyncratic risk in a given portfolio. On the other hand, such risk factors are typically systemic (macroeconomic) or defined for finer granularity economic clusters, such as industry and region sectors. They capture industrial, technology or government policy trends, providing a transparent and tractable link between macroeconomic conditions and credit risk outcomes. 

The latter approach is typically espoused by climate scenario analysis, especially when quantifying transition (i.e. policy) risk. This was the choice for the standardized climate scenario exercise (SCSE) \cite{scse_rel}, conducted in 2024 by the Office of the Superintendent of Financial Institutions of Canada (OSFI) and Qu\'{e}bec’s Autorit\'{e} des March\'{e}s Financiers (AMF). 

This paper presents the theoretical description of the stress-testing methodology based on the IFRS9/CECL framework. Its application is illustrated using the SCSE as a case study with individual examples being drawn from the SCSE’s official release and report webpages \cite{scse_rel, scse_rep}. Section~\ref{s:execution} outlines the methodology for assessing the impact of a scenario through the provisioning infrastructure when the effect of the scenario on the probability of default of an exposure is known. The design of a standardized scenario analysis under this framework requires a set of adjustment factors for the standardized exposure groupings. Section~\ref{s:methodology} then describes an approach for generalizing probability-of-default changes when the impact of a scenario is observed for a sample of exposures with known features.

\noindent{\bf Authors' note:} The authors were the principal designers of the relevant modules of the SCSE. However, the views expressed in this paper are the authors’ own and do not necessarily reflect the views of the Office of the Superintendent of Financial Institutions (OSFI), Autorit\'{e} des March\'{e}s Financiers (AMF), their current, or past employers. The authors did not conduct further investigations into the exemplified case studies; the examples are cited to the extent covered by the SCSE publications \cite{scse_rel, scse_rep}.

The authors would like to thank their colleagues for their valuable contributions. In particular, we are grateful to Brett Lindsey and Stevan Manokaran. The authors also thank Omneia Ismail for her review of the manuscript and for her insightful and constructive comments.

\section{Scenario Analysis Leveraging the Provisioning Mechanism}\label{s:execution}

In this section, we present a methodology that leverages provisioning mechanisms to assess the impact of underlying scenarios on each exposure accounted for under IFRS 9 or CECL, assuming the existence of an operator that translates baseline annual PDs into scenario-encoded annual PDs. Although such PD transformations may exhibit different structures and features, for the purposes of this section we assume a generic operator of this form. In Subsection~\ref{ss:pdAdjust} however, we propose a specific implementation of this operator that augments the baseline PD with a logarithmic add-on.

\subsection{Probability of Default Approach} 

We start by briefly introducing the probability-of-default approach to assessing the expected loss of exposures for provisioning. For a more detailed discussion of this approach, including its methodology and exceptions, we refer the reader to \cite{bellini}.

The probability-of-default-based approach to estimating credit losses is the primary methodology used to calculate provisions under both IFRS~9 expected credit loss (ECL) and CECL frameworks. Under this approach, expected credit losses are calculated using the general formulation
\[
ECL = \mathbb{E}( PD \times LGD \times EAD)
\]
where $\mathbb{E}$ denotes the expectation operator over lifetime of the exposure, $PD$ is the probability of default, $LGD$ loss given default, and $EAD$ exposure at default. For a finite set of scenarios and annual observations of credit outcomes the formula becomes 
\begin{equation}\label{eq:ecl_bs}
ECL = \sum_{k=1}^m w_k \sum_{t=1}^n \rm{PV} \left( pd_{k}^{(t)} \times lgd_{k}^{(t)} \times  ead_{k}^{(t)} \right)
\end{equation}
where for each macroeconomic scenario $k$:
\begin{itemize}
    \item $pd_{k}^{(t)}$, $lgd_{k}^{(t)}$, and $ead_{k}^{(t)}$ are  respectively PD, LGD and EAD of the exposure for year $t$,
    \item $\rm{PV}$ is the present value operator,
    \item $m$ is the number of underlying ECL scenarios used  to determine ECL of the exposure,
    \item $n$ is the remaining maturity of the exposure.
\end{itemize}
While the FASB’s CECL standard adopts lifetime expected loss perspective, the IASB’s IFRS~9 classifies exposures into three stages based on changes in credit risk, with only 12-month expected credit losses recognized for exposures that have not experienced a significant increase in credit risk. For the purpose of scenario analysis, we propose disregarding this staging mechanism and assessing the impact of scenarios over the lifetime of the exposures. 

\subsection{Scenario Analysis Using ECL}\label{ss:ecl}

When executing scenario analysis through narrative-driven PDs, both baseline and scenario-adjusted expected loss calculations can leverage the probability-of-default approach of the ECL accounting framework. Baseline ECLs are estimated using the existing lifetime ECL formulation for each exposure, as shown in the equation (\ref{eq:ecl_bs}), while scenario impacts are measured by adjusting the PD and LGD components in accordance with the underlying scenario.

The scenario-adjusted ECL ($ECL_{sc}$) is calculated in the same manner as the baseline ECL ($ECL_{bs}$), with baseline PDs and LGDs replaced by scenario-adjusted probabilities of default ($pd_{sc}$) and loss-given-default parameters ($lgd_{sc}$), respectively. Therefore, 
\begin{equation}\label{eq:ecl_sc}
ECL_{sc} = \sum_{k=1}^m w_k \sum_{t=1}^n \rm{PV} \left( pd_{sc,k}^{(t)} \times lgd_{sc,k}^{(t)} \times  ead_{k}^{(t)} \right).
\end{equation}
While the EAD component may also be affected by scenarios, for simplicity we assume that exposures at default are agnostic to the underlying scenarios.

The difference between the scenario-adjusted ECL and the baseline ECL, denoted by $\Delta ECL := ECL_{sc} - ECL_{bs}$, represents the impact of the underlying  scenario. 

\begin{adjustwidth}{.8cm}{0cm}
{\it \bf Case study:} 
For the SCSE, the instructions and quantitative requirements for calculating scenario-adjusted PDs were provided for each underlying climate scenario for all years from 2025 to 2050. While the baseline ECL corresponds to the lifetime expected losses of each exposure calculated under the IFRS~9 ECL framework, the climate-adjusted ECL calculations are performed at five-year intervals over the scenario projection horizon—namely 2030, 2035, 2040, and 2045—for each climate scenario within the scope of the exercise.
\end{adjustwidth}

In the following subsections, we discuss the methodology to estimate scenario adjusted PDs and LGDs. 

\subsection{Scenario Adjusted PDs}\label{ss:pd}

To reflect the common use of annual conditional PDs for provisioning, we assume that for each year $t$ in the scenario time horizon, climate-adjusted annual PDs at year $t$ are derived from baseline PDs, with both PDs  conditional on the exposure not having defaulted in prior periods (i.e., years $1,\ldots,t-1$).

 However, one should note that, the PDs used in the ECL formulation (\ref{eq:ecl_sc}) are unconditional. Therefore, to ensure consistency, unconditional PDs must be decomposed into annual conditional PDs ($cpd$).
To do so, we rely on this simple decomposition of 
\[
pd^{(t)} = ps^{(1,...,t-1)} \times cpd^{(t)}
\]
where $ps^{(1,...,t-1)}$ denotes the probability of survival of the exposure through the interval of years $1$ to $k-1$ while default probability in year $k$ is  $cpd^{(t)}$. 

While termination or impairment of an exposure may arise for several reasons - such as prepayment or restructuring, etc. - we adopt the following relationship between conditional and unconditional PDs:
\[
ps^{(1,...,t-1)} =\prod_{j=1}^{t-1} (1-cpd^{(j)})
\]
which is based on the simplifying assumption that an exposure can terminate prematurely only through default.

Accordingly, to derive conditional PDs for each year $t$, the following recursive formulas are used to derive the condition PDs from the baseline PDs $(pd_{bs}^{(t)})_t$:
\begin{align}
cpd^{(1)}_{bs} &=pd^{(1)}_{bs} \\
cpd^{(t)}_{bs} & = \frac{pd^{(t)}_{bs}}{\prod{j=1}^{t-1} (1-cpd^{(j)}_{bs})}
\end{align}
As mentioned above, we assume the scenario adjusted conditional PDs ($pd_{sc}^{(t)}$) are attained through an operator $\mathbb{P}_{t, sc}$, where 
\begin{equation}\label{eq:P-operator}
\mathbb{P}_{t,sc}(cpd^{(t)}_{bs} ) = cpd^{(t)}_{sc}. 
\end{equation} 
In Subsection~\ref{ss:pdAdjust}, we discuss a proposed construction of the operator $\mathbb{P}$ that overlays the baseline PD with a logarithmic add-on.

Similarly, unconditional scenario-adjusted PDs, $(pd_{sc}^{(t)})_t$, can be derived based on the same logic applied to the survival factors:
\[
pd_{(sc)}^{(t)} = \prod_{j=1}^{t-1} (1-cpd_{sc}^{(j)})\times cpd_{sc}^{(t)}.
\]

\subsection{Scenario adjusted LGDs}\label{ss:lgd}

Due to observable nature of defaults, the probability of default is generally easier to model then loss given default, which does not offer the same transparency. Consequently, assessing the impact of a scenario on PD is typically more straightforward. This is largely because the risk drivers for loss given default (LGD) often incorporate more asset-specific, idiosyncratic factors, which are harder to capture in high-level scenario analyses. However, assuming a scenario-agnostic LGD is not necessarily a sound assumption, as it may underestimate the impact of the scenario. For this reason, we propose to approximate LGD adjustments using the information available for PD.

The Frye-Jacobs relationship describes the interaction between changes in PD and LGD in credit risk modeling under certain simplifying assumptions \cite{frye}. It provides a framework for understanding how shifts in credit quality or macroeconomic conditions can simultaneously affect both the likelihood of default and the expected severity of loss given default. By linking PD and LGD, this relationship allows one to approximate either LGD or PD if the change in the other is known. However, this transformation should be used only as a proxy; when estimates for both PD and LGD are available, it is advisable to rely on the direct estimates rather than the approximation.

In the absence of direct estimates of scenario impacts on LGD, scenario-adjusted LGDs, $\left(lgd_{sc}^{(t)}\right)_t$, may be proxied using Frye–Jacobs relationship, based on the deviations observed in the PD component for each year $t$ over the scenario horizon. In this case, for each year $t$, the scenario-adjusted LGD is calculated as follows: 
\[
lgd_{sc}^{(t)} = \frac{\phi\left( \phi^{-1}(pd_{sc}^{(t)}) - \phi^{-1} (pd_{bs}^{(t)})+ \phi^{-1}( pd_{bs}^{(t)}\times lgd_{bs}^{(t)})\right)}{pd_{sc}^{(t)}}
\]
Where $\phi$ denotes the cumulative distribution function of the standard normal distribution, and $\phi^{-1}$ is its inverse. 

\subsection{Limitations}

\begin{itemize}
    \item To adjust survival factors within lifetime PD estimates, the methodology assumes that an exposure terminates only at contractual maturity or upon default. Other forms of early termination, such as prepayment, refinancing, or restructuring, are not explicitly modeled. This simplifying assumption may lead to material inaccuracies for portfolios in which such forms of exposure termination are common.
    
   \item Scenario impacts on credit risk are incorporated through adjustments to the PD and LGD parameters. The exposure-at-default (EAD) component is assumed to remain invariant across scenarios, thereby abstracting from potential scenario-driven changes in utilization rates, drawdown behavior, or amortization profiles.
   
   \item When direct estimates of scenario impacts on LGD are unavailable, scenario-adjusted LGDs are proxied using Frye–Jacobs relationship, expressed as a function of scenario-adjusted PDs, baseline PDs, and baseline LGDs (Subsection~\ref{ss:lgd}). This approach inherits the underlying assumptions of the Frye–Jacobs framework, including reliance on a Vasicek-style distribution for PD and LGD dynamics. For tractability, correlation parameters across Vasicek distributions are assumed to be zero, which may limit the framework’s ability to capture joint stress effects between default likelihood and loss severity. These assumptions may be particularly restrictive when the underlying risk drivers of PD and LGD differ or affect the two components in asymmetric ways.

\end{itemize}

\section{Sectoral Level Adjustments}\label{s:methodology}

In this section, we present a design methodology to construct risk factor adjustments at the level of standardized groupings of exposures where the groupings are defined based on common features of the exposures. We call such adjustments sectoral. This methodology relies on the availability of the impacts of a scenario on the probabilities of default for a sample of exposures with common features. We outline the steps involved in designing the adjustment factors. 

Similarly to Section~\ref{s:execution}, each step is first presented in a general context and then illustrated using the SCSE as a case study. For brevity, we omit common steps in model development, such as data validation, exploratory data analysis, and standard testing procedures.

\subsection{Development Data}
In what follows, each exposure is treated as a distinct “entity” in the development dataset. These entities represent individual units for which probabilities of default are calibrated - such as a firm, a counterparty, or an obligor - each categorized by a set of observable features/attributes. To formalize the notation, assume the sample of entities to be indexed by the set $\mathbf{I}$. For each entity $i \in \mathbf{I}$, let $\{ X_i^{(j)} \}_{j \in \mathbf{V}}$ denote the set of its features. Further assume that, for each $i \in \mathbf{I}$, forecast probabilities of default (PD) are available over a set of future time horizons $\mathbf{T}$ under both a baseline scenario, denoted by $bs$, $\{ pd_{i,bs}^{(t)} \}_{t \in \mathbf{T}}$, and a target scenario, denoted by $sc$, $\{ pd_{i,sc}^{(t)} \}_{t \in \mathbf{T}}$. Indexing ensures a consistent way to track entity level features/attributes, which is necessary because the adjustment factor methodology is applied to individual exposures before aggregating probabilities of default across groups or portfolios.

\begin{adjustwidth}{.8cm}{0cm}
{\it \bf Case study:} In the case of the SCSE, the exercise was based on a sample of 36,000 publicly traded entities deemed representative of the entities that participating financial institutions (as commercial lenders) hold in their portfolios.

For each entity in the sample, the industry sector, jurisdiction, and initial creditworthiness were known. Using the mappings developed for the exercise, these attributes were classified into 25 industry classes, nine regions, and six credit quality buckets, respectively. In the context of climate transition scenarios, additional transition relevant features/attributes, such as emissions intensity were also available, which influenced the entity-level PDs.

A vendor data source used by the SCSE provided $\bf{entity\ level}$ forward-looking probabilities of default for the baseline scenario and for each of the climate scenarios covered by the exercise. These vendor-supplied climate-adjusted probabilities of default were derived using three integrated assessment models (IAMs), namely GCAM, REMIND, and EPPA. We refer the reader to \cite{iam} for further details on IAMs.
\end{adjustwidth}

\subsection{Development Variables} 
To construct credit adjustments, we begin by using  a model that translates the narratives of the underlying scenarios into credit adjustments based on a set of independent variables $\mathbf{V}$, where these variables represent the observable features of each entity. Within this set, we distinguish a subset $\mathbf{V}_c \in \mathbf{V}$ consisting of the features along which entities are assumed to exhibit homogeneous impacts to the scenario. We also define a complimentary set, $\mathbf{V}_d := \mathbf{V}$, for the remaining independent variables which capture sources of heterogeneity. Suppose that, for the subset of variables $\mathbf{V}_c \in \mathbf{V}$ and for each time snapshot, the pattern of probabilities of default for entities within each bucket defined by $(j = J_0^{(j)})_{j \in \mathbf{V}}$ can be captured by such a model, conditional on the remaining variables in $\mathbf{V}_d$. Let $\mathbf{I}_0$ denote the subset of entities satisfying these conditions.

Then, for each snapshot $t \in \mathbf{T}$, the differences observed in $\{ pd_{i,bs}^{(t)} ,pd_{i,sc}^{(t)}  \}_{i \in \mathbf{I}_0}$ can be generalized into credit adjustments applicable to all entities that meet the full set of conditions $(j = J_0^{(j)})_{j \in \mathbf{V}}$.

\begin{adjustwidth}{.8cm}{0cm}
{\it \bf Case study:} In the case of the SCSE, it was assumed that all entities belonging to the same industry sector respond similarly to the climate scenarios; that is, $\mathbf{V}_c =\{\rm{industry}\}$. Accordingly, the SCSE industry mapping was designed to support this assumption, drawing primarily on the narratives and economic features of the underlying scenarios and the integrated assessment models. 

For example, it was assumed that all entities active in natural gas extraction exhibit changes in their probabilities of default in response to a given climate scenario that can be modelled conditional on their other features. Importantly, this assumption does not imply that all entities active in natural gas extraction behave identically under a climate scenario. Rather, it implies existence of a model—of the type described above—that can estimate entity-specific credit adjustments based on other fearures, such as region and credit risk bucket, represented by $\mathbf{V}_d$.
\end{adjustwidth}

\subsection{Sectoral Calibration}\label{ss:creditModels}

In this subsection, we discuss construction of $\bf{sectoral\ level}$ scenario PD adjustments using a logarithmic adjustment. 

To derive the risk drivers, for each bucket defined by $(j = J_0^{(j)})_{j \in \mathbf{V}_c}$,  each climate scenario $sc$, and each snapshot $t_0 \in \mathbf{T}$, we employ a linear regression model to estimate the difference in probabilities of default expressed in logit terms, where  the logit function is defined as $\rm{logit}(x) = \ln(\frac{x}{1-x})$.  Specifically, we model this difference using the linear relationship
\[
Y = f_{t_0,sc,(j = j_0^{(j)})_{j \in \mathbf{V}_c}}(X_v)_{v\in \mathbf{V}_d} := \sum_{v \in \mathbf{V}_d} \beta_v X_v + \alpha
\]
which is calibrated on the independent variables $\{ X_i^{(j)} \}_{j \in \mathbf{V}_d}$ and the dependent variable $\{ \rm{logit}(pd_{i,sc}^{(t_0)}) - \rm{logit}(pd_{i,bs}^{(t_0)}) \}_{i \in \mathbf{I}_0}$, yielding coefficients $\beta_v$ and constant $\alpha$.  The calibration is performed on the subset $\mathbf{I}_0 \subseteq \mathbf{I}$ consisting of all entities that fall within the bucket  $(j = J_0^{(j)})_{j \in \mathbf{V}_c}$.
Subsequently, 
\[
\delta_{t_0,sc} := f_{t_0,sc,(j = j_0^{(j)})_{j \in \mathbf{V}_c}}(J_0^{(j)})_{v\in \mathbf{V}_d}.
\]
which defines the logarithmic PD adjustment capturing the narrative of the underlying scenario for the designated bucket of each obligor.

\begin{adjustwidth}{.8cm}{0cm}
{\it \bf Case study:} 
For the SCSE, the dependent variables consisted of the current credit risk bucket, $X_r$, encoded as an ordinal variable, and a set of regional dummy variables $(X_k)_k$. This choice of variables was made to preserve the rank ordering of credit risk buckets while the inclusion of regional dummy variables proved particularly useful in calibrating models capable of generating meaningful regional narratives even in cases where the number of entities within a given bucket was insufficient to support robust statistical inference. In such cases, the calibrated models were still able to estimate adjustment narratives that were closer to, or further from, the curve implied by the credit risk bucket alone.

By contrast, in regions with a sufficiently large presence of entities within a given industry sector, the regional coefficients produced more distinct outcomes, better reflecting the features of those subsamples.

\end{adjustwidth}

\subsection{PD adjustments}\label{ss:pdAdjust}

Using the scenario adjustments factors  $(\delta_{t,sc})$ developed in Subsection~\ref{ss:creditModels}, the baseline PDs can be adjusted to obtain the PDs that capture the implications of the underlying scenarios.  Specifically, for each year $t$, if the baseline probability of default of an obligor is $pd_{t,bs}$, the projected PD of the obligor under  scenario $sc$, denoted by $pd_{t,sc} $, is obtained using the operator $\mathbb{P}_{sc,t}$ defined by 
\begin{equation}\label{eq:pd_overlay}
\mathbb{P}_{t,sc}(pd_{t,bs}) :=  \frac{1}{(1+\exp(-(logit(pd_{t,bs})+\delta_{t,sc}))}.
\end{equation}
This operator  overlays the logit  of the baseline PD with a logarithmic add-on developed from the underlying sample representing the relevant exposures.

\begin{adjustwidth}{.8cm}{0cm}
{\it \bf Case study:} 
For the SCSE, PD adjustments were constructed as described above for each underlying climate scenario across all combinations of industry sector, region, credit quality bucket, and each year over the scenario horizon.
\end{adjustwidth}

\subsection{Statistical Testing of the Sectoral Risk Factors} 
Since the underlying models are linear regressions calibrated on non-binary data—namely, logit differences of probabilities of default—standard diagnostic tests can be applied to assess their suitability for the intended purpose. These tests include, but are not limited to, the coefficient of determination ($R^2$), the F-test for overall model significance, and coefficient-level $t$-statistics and associated $p$-values.

Any unsatisfactory test results should be examined and interpreted carefully. In particular, insufficient granularity in the classification scheme—that is, a choice of $\mathbf{V}_c$ that is not sufficiently representative—may lead to poor model performance. In such cases, model developers should consider revisiting the partitioning of variables, $\mathbf{V} = \mathbf{V}_c \cup \mathbf{V}_d$, to improve model fit and interpretability.

As a general rule of thumb, we strongly encourage thorough exploratory data analysis, careful sample checks, and the use of abundant visualizations to ensure that the resulting models produce intuitive and robust results.

\subsection{Limitations}

\begin{itemize}
    \item {Note that the proposed construction uses as it inputs outputs of upstream models for baseline and scenario-adjusted probabilities of default. Any errors, limitations, or simplifying assumptions in these upstream models are carried forward into the proposed process.}
    
    \item The representativeness of the underlying sample can also introduce biases into the results. As noted above, some classes may be underrepresented in the sample; in such cases, the model effectively produces predictions that largely reflect the fitted curve for the entire sample.
    
    \item Lastly, the assumption that the same set of variables can produce a meaningful model across all segmentations is a simplifying assumption\footnote{In other words, the choice of $\mathbf{V}_c$ and $\mathbf{V}_d$ is assumed to be static across the entire sample. This is a notable simplifying assumption.} that could lead to model error, particularly in cases where different segments of the underlying sample respond differently to the scenarios.

\end{itemize}

\section{Conclusion}\label{s:conclusion}
Is it a novel idea to use IFRS 9 for scenario analysis? The conceptual differences are well known: IFRS 9 and stress testing serve distinct objectives. The former aims to generate expected credit losses for provisioning, using probability‑weighted, unbiased point‑in‑time (PiT) PDs and forward‑looking scenarios. The latter focuses on assessing capital adequacy under severe shocks, relying on adverse PiT PDs derived from hypothetical, often regulator‑mandated scenarios. While the idea of linking the two frameworks may not be entirely new and may have been explored by financial institutions, the specific implementation choices required to integrate them have not been previously explicated in the literature. The methodology we present is scenario‑agnostic, allowing PDs generated under typical stress‑testing scenarios to be applied directly within the IFRS9/ECL framework.

Sectoral risk factors become particularly important when expected credit losses cannot be reliably measured using only macro‑level variables. In situations where the impact of a shock varies according to common features of the exposures, sectoral differentiation is essential for producing more robust and precise loss estimates. For example, a standard macro stress test or IFRS 9 ECL framework typically relies on a set of macroeconomic variables that do not vary across sectors of the economy, which limits their ability to capture heterogenous impacts. This stands in contrasts to scenarios such as the introduction of tariffs or carbon‑pricing policies, where the impacts differ markedly by sector. In such cases, incorporating sector‑specific risk factors is critical to accurately capturing the distribution of losses across the portfolio.

The aggregation from entity to sectoral level via a statistical pattern discovery techniques such as regression is one way to achieve the desired risk factor granularity, provided the input is available at the entity level and is sufficient and representative for each target segment. Proxying and other missing data treatments may be used for select underrepresented target sectors. 

In conclusion, the proposed approach is flexible across scenario designs, allowing stress‑testing PDs to feed directly into IFRS9 based loss estimates. By incorporating sector‑specific risk drivers and aggregating entity‑level effects into coherent sectoral impacts, the methodology offers a consistent and scalable framework for capturing credit losses across a wide range of economic conditions.

\end{document}